\documentstyle[amssymb,prl,aps]{revtex}

\begin{document}
\title{Complete Bell state measurement with controlled photon absorption and
quantum interference}
\author{Akihisa Tomita}
\address{Fundamentral Research Laboratories, System Devices and Fundamental Research,%
\\
NEC Corporation, 34 Miyukigaoka, Tsukuba, Ibaraki 305-8501, Japan}
\date{Received date }
\maketitle

\begin{abstract}
A solid state device to discriminate all the four Bell states is proposed.
The device is composed of controlled absorption crystals, rotators, and
retarders. The controlled absorption, where the state of one photon affects
the absorption of the other photon, is realized by two photon absorption in
a cubic crystal. The controlled absorption crystal detects a particular Bell state
and is transparent for the other Bell states. The rotators and retarders
transform a Bell state to another. This device may solve the problems in the
early quantum teleportation experiments in photon polarization states.
\end{abstract}


The quantum information technology aims to achieve performance in
communication and computation systems superior to those based on classical
physics by utilizing the entanglement between particles. Bell states, the
maximally entangled two-particle states, are crucial ingredients in the
quantum information technology, such as quantum teleportation\cite{Bennett93}%
, dense coding\cite{Bennett92}, and entanglement swapping\cite{Zukowski93}.
The quantum teleportation transfers unknown quantum states. Dense coding
offers larger capacity than the classical communication. The entanglement
swapping creates entanglement of two particles that have never interacted,
and provides multi-particle entanglement. Quantum repeaters\cite{Briegel98},
which are indispensable elements in quantum networks to relay unknown
quantum states in a long distance, can be realized with the quantum
teleportation and the entanglement swapping. The key devices for the above
quantum information processing are generators and discriminators of the four
Bell states:
\begin{eqnarray}
\left| \Phi ^{\left( \pm \right) }\right\rangle  &=&\frac{1}{\sqrt{2}}\left(
\left| x\right\rangle _{1}\left| x\right\rangle _{2}\pm \left|
y\right\rangle _{1}\left| y\right\rangle _{2}\right)   \label{Bell} \\
\left| \Psi ^{\left( \pm \right) }\right\rangle  &=&\frac{1}{\sqrt{2}}\left(
\left| x\right\rangle _{1}\left| y\right\rangle _{2}\pm \left|
y\right\rangle _{1}\left| x\right\rangle _{2}\right) .  \nonumber
\end{eqnarray}
A parametric downconversion in a nonlinear crystal generates one of the
two-photon Bell states. The other three Bell states are easily obtained by
suitable unitary operations with linear polarization elements (such as
retarders and rotators\cite{Mattle96}.) On the other hand, a realizable
complete Bell state measurement (BSM) method, which discriminates all the
four Bell states, has been little known. L\"{u}tkenhaus {\it et al}\cite
{Lutkenhaus99} have proved a no-go theorem that shows the impossibility of
never failing BSM with linear optical elements and detectors. The lack of
the complete BSM limits the universality of the quantum teleportation
experiments. Bouwmeester {\it et al}\cite{Bouwmeetster}{\it \ }has succeeded
the quantum teleportation on the polarization states in only 25 \% of the
input states, because they discriminated only one ($\Psi ^{\left( -\right) }$%
) state out of the four Bell states. Boschi {\it et al}\cite{Boshi98}
reported 100 \% efficiency with linear optical elements by implementing the
Bell states on the product space of the two degrees freedom of \ the same
system. The teleported state had to be prepared in the system and was known
before the teleportation. A controlled NOT gate transforms the Bell states
into disentangled states to be easily discriminated, and provides a quantum
circuit for the complete BSM\cite{Bruss97}. However, a controlled NOT gate
itself is beyond current technology.

Recently, a progress towards a realizable BSM has been made. Scully {\it et
al}\cite{Scully99} proposed a two-photon absorption (TPA) scheme for the
complete BSM. Unlike the quantum gates, coherency is not required in the
output of the TPA detection scheme; the measurement is done in the TPA
process. Therefore, the photon energy can be resonant to the atomic
two-photon transition. This resonance may enhance the TPA to resolve the low
efficiency problem of the nonlinear crystals\cite{Bennett92}. The TPA scheme
\cite{Scully99}, however, requires three gas cells filled with atoms
prepared in special coherent superpositions of hyperfine states to
discriminate the Bell states. It is still not easy to prepare the coherent
superposition; it requires other strong laser sources with carefully tuned
frequencies and phases. In this letter we propose more realizable devices
for the complete BSM. We will show that a particular Bell state can be
discriminated without any state preparation, and that the discrimination of
only one Bell state is enough for the complete BSM with the help of linear
polarization elements that transform one Bell state to another.

The complete BSM can be achieved as follows. We found that a controlled
absorption and quantum interference result in the discrimination of a
particular Bell state. The ``controlled absorption'' refers to the phenomena
that the state of one photon affects the absorption of the other photon. An
example of the controlled absorption is TPA in crystals. The TPA coefficient
in a crystal depends on the polarizations of the two photons (two-photon
polarization selection rule.) The quantum interference will take place if
the transition to the same final state is possible through two routes. The
following example illustrates the Bell state discrimination. Suppose two
photons in one of the four Bell state enter a crystal. The Bell states are
given by Eq.(\ref{Bell}) in the linear polarization bases: $\left|
x\right\rangle $ $\left( \left| y\right\rangle \right) $ denotes the photon
polarized in $x$ ($y$) direction, as in the Innsbruck experiments\cite
{Bouwmeetster}. If the two-photon polarization selection rule in the crystal
allows the TPA with the photons only in the same polarizations, i.e., $%
\left| x\right\rangle _{1}\left| x\right\rangle _{2}$ and $\left|
y\right\rangle _{1}\left| y\right\rangle _{2}$, absorption of the photons in
the $\Psi ^{\left( \pm \right) }$ states is forbidden. Moreover, if the two
combinations of photons $\left| x\right\rangle _{1}\left| x\right\rangle _{2}
$ and $\left| y\right\rangle _{1}\left| y\right\rangle _{2}$ contribute to
the transition to the same final state with the same amount, the quantum
interference between the transitions by $\left| x\right\rangle _{1}\left|
x\right\rangle _{2}$ and $\left| y\right\rangle _{1}\left| y\right\rangle
_{2}$ photons enhances the TPA of the $\Phi ^{\left( +\right) }$state but
suppresses that of the $\Phi ^{\left( -\right) }$ state. Detection of TPA by
two-photon luminescence or two-photon photoconductivity will thus indicate
that $\Phi ^{\left( +\right) }$ state entered the crystal. The TPA from the $%
\Gamma _{1}^{+}$ ground state to the $\Gamma _{1}^{+}$ final state in a
cubic crystal obeys the above two-photon polarization selection rule and
shows the quantum interference, as shown later. The crystal is transparent
to the photons in the other Bell states. The other Bell states can be
discriminated by the crystal after the state transformation to the $\Phi
^{\left( +\right) }$ state. A $\pi /2$-retarders (quarter-wave plates)
applied to both photons produce a phase difference $\pi /2$ between $x$%
-polarized photons and $y$-polarized photons\cite{Stenholm96}, and thus
transform the $\Phi ^{\left( -\right) }$ state to the $\Phi ^{\left(
+\right) }$ state, but leave the $\Psi ^{\left( \pm \right) }$ states
unchanged except a common phase factor. Rotating the polarization of one
photon by $\pi /2$ interchanges the $\Psi ^{\left( \pm \right) }$ states and
the $\Phi ^{\left( \mp \right) }$ states. \ref{figone} depicts a complete
BSM device composed of the TPA\ crystals, linear polarization elements, and
detectors. The crystals absorb only the photons in the $\Phi ^{\left(
+\right) }$ state. The $\Phi ^{\left( +\right) }$ state is discriminated by
the first crystal. The transmitted states are transformed by the $\pi /2$%
-retarders. Then the second crystal discriminates the $\Phi ^{\left(
-\right) }$ state. The photons transmitted the second crystal must be in the
$\Psi ^{\left( \pm \right) }$ states, which are transformed to the $\Phi
^{\left( \mp \right) }$ states by the rotator. The third crystal thus
discriminates the $\Psi ^{\left( -\right) }$ state. Putting $\pi /2$%
-retarders to the transmitted beams, we can detect the $\Psi ^{\left(
+\right) }$ state by the forth crystal . All the four Bell states are
successfully discriminated. If the TPA crystal discriminates another Bell
state by different two-photon polarization selection rule, complete BSM can
be achieved similarly but with a different combination of retarders and
rotators.  It seems that the complete BSM may be possible with a device
composed of three TPA crystals that discriminate three Bell states, and an
ordinary photo-detector that detects the photon transmission from the three
TPA crystals. If the photons are detected in the photo-detector, one may
conclude the photons are in the forth Bell state. It is not always true,
however, because the TPA probability is less than unity. The photo-detector
would detect photons that the TPA crystals failed to absorb. The failure in
TPA discrimination will results in the error in BSM.

The two-photon polarization selection rule is derived from the second order
perturbation theory\cite{Inoue65}. We will consider TPA in a cubic crystal
belonging to the point group{\it \ O}$_{h}$, for simplicity. The crystal
ground state is assumed to belong to the $\Gamma _{1}^{+}$ irreducible
representation of energy $E_{0}$, the final state $\left| f_{m}^{\mu
}\right\rangle $ to the $m$-th row of the irreducible representation $D^{\mu
}$ of energy $E_{f^{\mu }}$. The total momentum vector ${\bf P}$ belongs to
the irreducible representation $D^{\lambda }=\Gamma _{4}^{-}$, and the final
state should belong to one of the decomposition of the product
representations as $\Gamma _{4}^{-}\otimes \Gamma _{4}^{-}=\Gamma
_{1}^{+}\oplus \Gamma _{3}^{+}\oplus \Gamma _{4}^{+}\oplus \Gamma _{5}^{+}.$
The Wigner-Eckart theorem separate the reduced matrix elements and the
geometrical factor of the TPA rate of two photons with the energies $\hbar
\omega _{1}$ and $\hbar \omega _{2}$\cite{Hamermesh62}\cite{Doni74}:
\begin{equation}
\alpha \left( \omega _{1},\omega _{2}\right) \varpropto
\mathrel{\mathop{\sum }\limits_{f^{\mu }}}%
\left|
\mathrel{\mathop{\sum }\limits_{m}}%
G_{\mu m}\left( \psi \right)
\mathrel{\mathop{\sum }\limits_{\varphi ^{\lambda }}}%
\Lambda _{\varphi ^{\lambda }}^{\pm }\left\langle f^{\mu }\left\| P^{\lambda
}\right\| \varphi ^{\lambda }\right\rangle \left\langle \varphi ^{\lambda
}\left\| P^{\lambda }\right\| 0\right\rangle \right| ^{2}\delta \left(
E_{f^{\mu }}-E_{0}-\hbar \omega _{1}-\hbar \omega _{2}\right) ,  \label{TPA2}
\end{equation}
where the energy denominator is defined by
\begin{equation}
\Lambda _{\varphi ^{\lambda }}^{\pm }=\frac{1}{E_{\varphi ^{\lambda
}}-E_{0}-\hbar \omega _{1}}\pm \frac{1}{E_{\varphi ^{\lambda }}-E_{0}-\hbar
\omega _{2}}.  \label{denom}
\end{equation}
The positive (negative) sign in the energy denominator (\ref{denom}) holds
for the final states belonging to the symmetric (antisymmetric)
decomposition of $D^{\lambda }\otimes D^{\lambda }$. TPA to the states
belonging to the antisymmetric decomposition ($\Gamma _{4}^{+}$) is weak for
nearly degenerate photons $\omega _{1}\thicksim \omega _{2}$. We will focus
the final states belonging to the symmetric decomposition ($\Gamma _{1}^{+}$%
, $\Gamma _{3}^{+}$, and $\Gamma _{5}^{+}$). The two-photon polarization
selection rule is determined by the geometrical factor
\begin{equation}
G_{\mu m}\left( \psi \right) =%
\mathrel{\mathop{\sum }\limits_{ll^{\prime }}}%
\left\langle vac\right| a_{1}^{l}a_{2}^{l^{\prime }}\left| \psi
\right\rangle \left( \mu m|\lambda l,\lambda l^{\prime }\right) ,
\end{equation}
where $\left( \mu m|\lambda l,\lambda l^{\prime }\right) $ refers to the
Clebsh-Gordan coefficients. We have generalized the expression by Doni {\it %
et al}\cite{Doni74} to include the non-classical photon states $\left| \psi
\right\rangle $. The operator $a_{1}^{l}(a_{2}^{l^{\prime }})$ is the $%
l(l^{\prime })$-th raw of the annihilation operators of photon 1(2). The
Clebsh-Gordan coefficients for the symmetric decomposition of the product
representation can be found in Ref. \cite{Koster63}. If the final state
belongs to the $\Gamma _{1}^{+}$ irreducible representation, the
Clebsh-Gordan coefficients read $\left( \Gamma _{1}^{+}|\Gamma
_{4}^{-}l,\Gamma _{4}^{-}l^{\prime }\right) =\left( 1/\sqrt{3}\right) \delta
_{l,l^{\prime }}.$ The geometrical factor
\begin{equation}
G_{\Gamma _{1}^{+}}\left( \psi \right) =\frac{1}{\sqrt{3}}\left\langle
vac\right| a_{1}^{x}a_{2}^{x}\left| \psi \right\rangle +\frac{1}{\sqrt{3}}%
\left\langle vac\right| a_{1}^{y}a_{2}^{y}\left| \psi \right\rangle +\frac{1%
}{\sqrt{3}}\left\langle vac\right| a_{1}^{z}a_{2}^{z}\left| \psi
\right\rangle   \label{geometric}
\end{equation}
takes $2/\sqrt{3}$ for the $\Phi ^{\left( +\right) }$ state and 0 for the $%
\Phi ^{\left( -\right) }$ state and the $\Psi ^{\left( \pm \right) }$
states, i.e.,   TPA occurs for only $\left| \Phi ^{\left( +\right) }\right) $
state in $\Gamma _{1}^{+}\rightarrow \Gamma _{1}^{+}$ transition. TPA is
allowed for two photons in parallel polarization in this transition. The
interference between the TPA process of $xx$-polarized photons and $yy$%
-polarized photons cancels the geometrical factor for the $\Phi ^{\left(
-\right) }$ state. The selection rules for the transition to the final
states belonging to other irreducible representations can be derived
similarly. The same selection rules hold to the crystals belonging to other
cubic symmetry groups like {\it T}$_{{\it d}}$. \

The Bell state detection requires a large TPA coefficient $\beta $. A
promising candidate is the two-photon transition to the biexciton states.
The giant oscillator strength and the resonance of the intermediate state
(exciton states) result in huge enhancement in TPA coefficient, which was
estimated to be 10$^{7}$ in CuCl\cite{Hanamura73}. \ The biexcitons in CuCl
belongs to the $\Gamma _{1}$ irreducible representation of the {\it T}$_{%
{\it d}}$ point group. The TPA photon energy lies at 3.186 eV, well
separated from the exciton transition at 3.202 eV to avoid undesirable
one-photon absorption. A cavity will enhance the TPA coefficient to improve
the detection efficiency. We assume that two-photons of frequency $\omega $
are confined in a cavity of volume $V$ filled by a CuCl crystal. The
electric field in the cavity is given by $E=\left( \hbar \omega
/n^{2}\epsilon _{0}V\right) ^{1/2}$, where $n$ is the refractive index in
the cavity and $\epsilon _{0}$ is the permittivity of the vacuum. The TPA
rate $\alpha $ in the cavity reads
\begin{equation}
\alpha =\left( \frac{c}{n}\right) ^{2}\epsilon _{0}\beta E^{2}=\frac{%
c^{2}\beta \hbar \omega }{n^{4}V}.
\end{equation}
Putting the values $\hbar \omega =3.186$ eV, $n\thicksim 3$, $\beta
\thicksim 0.1$ cm/W, and $V\thicksim $1 $\mu $m$^{3}$, we obtain the value
of the TPA\ rate: $\alpha \thicksim 6\times 10^{11}$ s$^{-1}$. The condition
to obtain efficient TPA is that the photon life time of the cavity is longer
than $\alpha ^{-1}\thicksim $1.7 ps. This implies the $Q$ value of the
cavity to be larger than 8$\times 10^{3}$, which can be satisfied by the
current technology. The complete BSM by a solid state device is thus be
realizable in the present TPA detection scheme.

BSM with TPA has a further advantage. We can set the photon energy of one
beam to be different from that of the other beam, and the sum of the
energies to be resonant to the two photon transition, i.e., $\hbar \omega
_{1}\neq \hbar \omega _{2}$ and $\hbar \omega _{1}+\hbar \omega
_{2}=E_{f^{\mu }}-E_{0}.$ This nondegenerate BSM can rule out the
possibility that detects the two photons from one source, because the
non-resonant TPA of the photons with the same energy will be weak. It will
remove the difficulty in the Innsbruck experiments\cite{Bouwmeetster}
pointed out by Braunstein and Kimble\cite{Braunstein98}.

The controlled absorption can be realized in other systems. We here consider
the optical transition between the lowest sublevels in a quantum dot. We
assume the lowest hole states are the heavy hole states $\left| 3/2,\pm
3/2\right\rangle _{h}$. Then the electron-hole pair states result from the
optical transition are $\left| \uparrow \right\rangle =\left|
1/2,-1/2\right\rangle _{e}\left| 3/2,-3/2\right\rangle _{h}$ and $\left|
\downarrow \right\rangle =\left| 1/2,1/2\right\rangle _{e}\left|
3/2,3/2\right\rangle _{h}$; the former is created by a right-handed
circularly polarized photons $\left| \sigma ^{+}\right\rangle $ and the
latter by a left-handed circularly polarized photons $\left| \sigma
^{-}\right\rangle $. Suppose one electron-hole pair exists in the quantum
dot, and define Bell states of an electron-hole pair and a photon:
\begin{eqnarray}
\left| \Phi ^{\left( \pm \right) }\right\rangle  &=&\frac{1}{\sqrt{2}}\left(
\left| \uparrow \right\rangle \left| \sigma ^{+}\right\rangle \pm \left|
\downarrow \right\rangle \left| \sigma ^{-}\right\rangle \right)
\label{QDBS} \\
\left| \Psi ^{\left( \pm \right) }\right\rangle  &=&\frac{1}{\sqrt{2}}\left(
\left| \uparrow \right\rangle \left| \sigma ^{-}\right\rangle \pm \left|
\downarrow \right\rangle \left| \sigma ^{+}\right\rangle \right)   \nonumber
\end{eqnarray}
These states are created \ when the quantum dot absorbs one of the two
photons in the Bell states (Eq. \ref{Bell}.) The state of two electron-hole
pairs should be in the form $\left( 1/\sqrt{2}\right) \left( \left| \uparrow
\right\rangle _{1}\left| \downarrow \right\rangle _{2}+\left| \downarrow
\right\rangle _{1}\left| \uparrow \right\rangle _{2}\right) $, so that only
the $\Psi ^{\left( +\right) }$ state in Eq. \ref{QDBS} is absorbed by the
quantum dot. This controlled absorption results from Pauli exclusion
principle. Linear polarization elements also transform the Bell states. A $%
\pi $-retarder, which transforms the $\left| \sigma ^{+}\right\rangle $
polarization state to the $\left| \sigma ^{-}\right\rangle $ state,
interchanges the $\Phi ^{\left( \pm \right) }$ states and the $\Psi ^{\left(
\pm \right) }$ states. A $\pi /2$-rotator, which provides relative phase
(-1) between the $\left| \sigma ^{+}\right\rangle $ polarization state and
the $\left| \sigma ^{-}\right\rangle $ state, exchange the signs as $\Phi
^{\left( \pm \right) }\rightarrow \Phi ^{\left( \mp \right) }$ and $\Psi
^{\left( \pm \right) }\rightarrow \Psi ^{\left( \mp \right) }$. The light
beam should go through the quantum dot four times, because the electron
stays in the excited quantum dot. The states are discriminated by the time
of the photon detection event. Therefore, the electron-hole state should
remain until the Bell state discrimination is completed. The time for the
Bell discrimination would be determined by the time resolution of the photon
detection. This requirement may limit the feasibility of the quantum dot BSM
devices.

The controlled absorption would be useful for other quantum circuit than the
complete BSM. The TPA scheme can be regarded as an integrated device of a
two-qubit quantum gate and a detector. It will provide a quantum circuit
where the qubits pass through quantum gates only once. Multi-photon
absorption process may realize more complicated quantum circuits, instead of
cascading discrete quantum gates.

\begin{figure}[tbp]
\caption{A proposed device for the complete Bell state measurement composed
of two-photon absorbing crystals (X), $\protect\pi /2$-retarders (quater-wave
plates), and $\protect\pi /2$-rotators. The crystals absorb the $\Phi
^{(+)} $ state, and the absorption is detected. The retaders and the rotators transform one Bell state to
another.}
\label{figone}
\end{figure}


\begin{references}
\bibitem{Bennett93}  C.H. Bennett {\it et al} , Phys. Rev. Lett., {\bf 70},
1895 (1993).

\bibitem{Bennett92}  C.H. Bennett and S.J. Wiesner, Phys. Rev. Lett. {\bf 69}%
, 2881 (1992).

\bibitem{Zukowski93}  M. Zukowski, A. Zeilinger, M.A. Horne, and A. Ekert,
Phys. Rev. Lett. {\bf 71}, 4287 (1993).

\bibitem{Briegel98}  H.-J. Briegel, W. Dr, J.I. Cirac, and P. Zoller, Phys.
Rev. Lett. {\bf 81}, 5932 (1998).

\bibitem{Mattle96}  K. Mattle {\it et al}, Phys. Rev. Lett. {\bf 76}, 4656
(1996).

\bibitem{Lutkenhaus99}  N. L\"{u}tkenhaus, J. Calsamiglia, and K.-A.
Suominen, Phys. Rev. {\bf A59}, 3295 (1999).

\bibitem{Bouwmeetster}  D. Bouwmeester {\it et al}, Nature {\bf 390}, 575
(1997).

\bibitem{Boshi98}  D. Boschi {\it et al}, Phys. Rev. Lett. {\bf 80}, 1121
(1998).

\bibitem{Bruss97}  D. Bruss {\it et al}, Phil. Trans. R. Soc. London {\bf %
A355}, 2259 (1997).

\bibitem{Scully99}  M.O. Scully, B.-G. Englert, and C.J. Bednar, Phys. Rev.
Lett. {\bf 83}, 4433 (1999).

\bibitem{Stenholm96}  S. Stenholm, Opt. Commun. {\bf 123}, 287 (1996).

\bibitem{Inoue65}  M. Inoue, and Y. Toyozawa, J. Phys. Soc. Jpn. {\bf 20},
363 (1965).

\bibitem{Hamermesh62}  M. Hamermesh, {\it Group Theory and Its Application
to Physical Problems}, (Addison-Wesley, Reading, 1962).

\bibitem{Doni74}  E. Doni, R. Girlanda, and G.P. Parravicini, Phys. Stat.
Sol. {\bf b65}, 203 (1974).

\bibitem{Koster63}  G.F. Koster, J.O. Dimmock, R.J. Wheeler, and H. Statz,
{\it Properties of the Thirty-Two Point Groups}, (MIT Press, Cambridge,
1963).

\bibitem{Hanamura73}  E. Hanamura, Solid State Commun. {\bf 12}, 951 (1973).

\bibitem{Braunstein98}  S.L. Braunstein and H.J. Kimble, Nature {\bf 394},
840 (1998).
\end{references}
\end{document}